\documentclass[epj]{svjour}
\usepackage{amsmath,amsfonts,amssymb}
\usepackage{color}
\usepackage{graphicx}
\usepackage{graphics}
\newcommand{\be}{\begin{eqnarray}}
\newcommand{\ee}{\end{eqnarray}}
\newcommand{\nn}{\nonumber}
\newcommand{\la}{\langle}
\newcommand{\ra}{\rangle}
\newcommand{\lla}{\langle\hspace{-.5mm}\langle}
\newcommand{\rra}{\rangle\hspace{-.5mm}\rangle}
\newcommand{\pa}{\partial}

\begin{document}
\title{Equation of state for agents on graphs}
\subtitle{}
\author{Arkadiusz Majka\inst{1} \and Wojciech Wi\`slicki\inst{1,2}
}                     
%
%
\institute{Interdisciplinary Centre for Mathematical and Computational Modelling, University of Warsaw, Pawi\`nskiego 5a, \\ PL-02-106~Warszawa \and 
A. So\l tan Institute for Nuclear Studies, Ho\.za 69, PL-00-681 Warszawa}
\date{Received: date / Revised version: date}
%
\abstract{
Choice models for populations of agents on graphs are studied in terms of statistical thermodynamics.
Equations of state are derived and discussed for different connectivity schemes, utility approximations, and temperature and volume regimes.
Analogies to ideal classical and quantum gases are found and features specific for network systems are discussed.
\PACS{
      {}{05.20.-y, 05.70.-a, 89.65.-s, 89.75.Hc}  
     } 
} 
\maketitle
\section{Introduction} 
Ensembles of active, choice-making individuals represent a natural field of extension for statistical physics. 
This broadening of theory's scope is interesting for a physicist because of a new quality brought to the study by autonomy of system's constituents in making decisions.
In addition, it is not {\it a priori} clear how to formulate an analog of the classical, equilibrium Boltzmann-Gibbs thermodynamics for systems where the notion of physical energy is not defined.
How probability measure on the state space is defined then?

Choice models, being part of decision theory, are widely applied to economics, psychology and social sciences.
For example, models used in demand analysis of transportation and communication usually assume that demand represents the result of decisions of individuals in the population.
These decisions consist of choices made among finite sets of possibilities.
Consider decision to be taken by potential passenger willing to travel between two nodes.
This client has to choose among offers of carriers, accounting for many factors relevant for decision, as e.g. the timetable and how it relates to his needs, durations of travels, fares, numbers of stops and changes on the route, probability of delays, declared and expected quality of service and many other aspects, too numerous to itemize.
In order to quantify such process of decision making, one incorporates discrete choice models in hope of better understanding and predicting behaviour of such complex system as transportation network and thus obtaining hints for marketing and revenue management.

Provided one knows how to formulate statistical thermodynamics for populations of agents, our understanding and predicting power for such systems is enhanced, as compared to choice models.
Having probability distributions in equilibrium one calculates thermodynamic potentials and response functions, and their evolutions with intensive variables as temperature, fugacity or pressure.
Moments of extensive variables, directly related to thermodynamic potentials, have clear economic value, e.g. market and route occupancies or fluctuations of numbers of clients on them.
In addition, monitoring collective phenomena and phase transitions is feasible using thermodynamic formalism.
Such tools are lacking in economy and social sciences.

In our previous papers we demonstrated how the system consisting of a set of decision makers on a network can be treated in the framework of equilibrium statistical thermodynamics and its generalization using R\'enyi entropies \cite{majka1,majka2}.
Applying thermodynamic formalism one tacitly assumes existence of stationary regime and states optimal with respect to entropy.
Finding thermal equilibrium is therefore related to the class of optimization problems, in particular to optimal path distribution on the network.
Problems of optimal path search in discrete systems, quite often met in condensed matter, lattice field theory, networks of molecular reactions, telecommunication and energy routing, were often treated non-thermodynamically (cf. e.g. refs \cite{burgos}).
We believe that application of Metropolis algorithms to this class of problems could be also effective.
Interesting for us, however, was not the optimization itself but studying global properties of the system in quasistatic approximation, assuming ergodic hypothesis.
For this we found thermodynamic approach to be well suited.

Building thermodynamic description, the central point was to find a probability measure on the event space.
We studied analogy between the utility function, used to quantify decision likelihood, and the total energy function describing probability of states in physics.
We found that for a properly defined system and utility function this analogy can be made exact.
In particular, this concerns the properties of additivity and extensivity of utility, and entails incorporation of the Boltzmann-Gibbs state probability measure, as for physical systems with no long-range correlations.

Building this analogy, however, we have not made it complete and never arrived to the equation of state for populations of agents on graphs. 
In this paper, after recapitulation of definitions of the system and statistical ensembles in chapter \ref{sec:1}, we introduce another extensive variable for the system, the market and network volume, and the conjugate intensive variable being an analogue of the pressure, and discuss equation of state of the network, both in general and for specific network topologies.
Analogies to behaviour of classical and quantum gases are highlighted, wherever found.

\label{intro}
\section{The system and the statistical ensembles}
\label{sec:1}

\begin{figure}[h]
\begin{center}
\includegraphics[scale=.5]{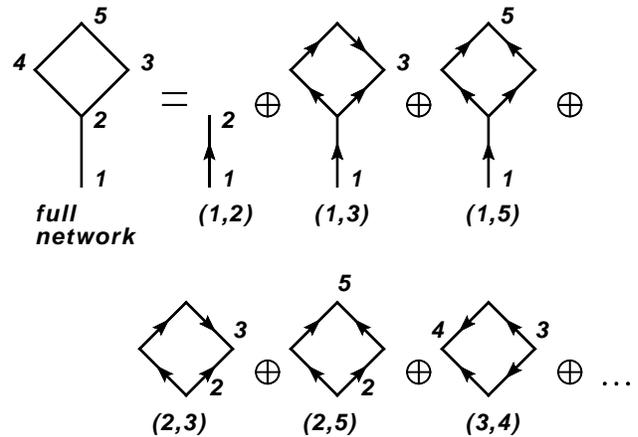}
\caption{\label{fig00}\em Decomposition of the communication network into markets, where only examples of representative markets are shown. The full list of markets and their decomposition into Hamilton routes is given in the text.}
\end{center}
\end{figure}

Our system $\mathcal S$ consists of a network $\mathcal G$, represented by a set of directed graphs, and a set of agents willing to find a directed route from the origin node $O$ to the destination node $D$. 
Each ordered pair $(OD)$ and set of all Hamilton routes\footnote{A Hamilton route is a path between two vertices of a graph that visits each vertex exactly once. Restriction to Hamilton routes is reasonable for applications to communication and transportation problems where loops should be avoided. Generalization of the formalism to all conceivable paths on a graph is straightforward, however we did not try to find analytic solutions in this case.} from $O$ to $D$ constitute, adopting economic terminology, the $k$-th {\it market} and is represented by a directed graph $\mathcal G_k$ ($k=1,\ldots,M$, $M$ denoting the total number of ordered $(OD)$ pairs in the network).
The whole network is the union of market graphs, $\mathcal{G}=\mathcal{G}_1\cup\ldots\cup \mathcal{G}_M$.
Market graphs can overlap, e.g. two subgraphs $\mathcal{G}_i$ and $\mathcal{G}_j$ can have common nodes or edges.
Such decomposition of the market graph $\mathcal{G}$ induces decomposition of the system $\mathcal S$ into the sum of markets $\mathcal{S}=\mathcal{S}_1\oplus\ldots\oplus\mathcal{S}_M$, where $\oplus$ stands for the sum of graphs {\it and} agents ascribed to corresponding graphs. 
Fig.~\ref{fig00} illustrates decomposition of an example system into markets.
Each market $(OD)$ can be further split into routes $[O,C_1,\ldots,C_L,D]$ in the same way, e.g. for the network from Fig.~\ref{fig00} decomposition is the following:
\be
(1,2) & = & [1,2] \nn \\
(1,3) & = & [1,2,3]\oplus [1,2,4,5,3] \nn \\
(1,4) & = & [1,2,4]\oplus [1,2,3,5,4] \nn \\
(1,5) & = & [1,2,3,5]\oplus [1,2,4,5] \nn \\
(2,3) & = & [2,3]\oplus [2,4,5,3] \nn \\
(2,4) & = & [2,4]\oplus [2,3,5,4] \nn \\
(2,5) & = & [2,3,5]\oplus [2,4,5] \nn \\
(3,4) & = & [3,2,4]\oplus [3,5,4] \nn \\
(3,5) & = & [3,5]\oplus [3,2,4,5] \nn \\
(4,5) & = & [4,5]\oplus [4,2,3,5] \nn \\
 & & \mbox{\small plus all reversed $(OD)\rightarrow (DO)$ pairs} \nn
\ee

The state of the system is defined by choices of routes performed by all agents for all markets. The state space consists of all possible distributions of all agents for all markets.

We consider so called {\it disaggregate model} where agents decide independently of each other. 
This assumption is not restrictive for a wide class of applications.
Notion of an agent can be often redefined and refer to groups of agents taking decisions as a whole, and only mutual independence of groups is relevant. 
Generally, the mechanism of group decision making can affect utility function.
We do not pretend to discuss rather vast subject of group decisions and refer to one of recent reviews \cite{ducheneaut}.

The utility $U_{i_k}$ of the $i_k$-th agent's choice ($i_k=1,\ldots,V_k$, $V_k$ being the number of choice alternatives on the $k$-th market and $k=1,\ldots,M$ the market index) is a finite, real random variable on the choice set and is defined axiomatically using the preference-indifference operator \cite{bierlaire1}. 
This operator relates any pair of alternatives from the choice set and is assumed to be reflexive and transitive, thus ordering the choice set linearly.
The choice set is assumed to be finite which guarantees existence of the best alternative.
These properties of the preference-indifference operator induce the linear order in the real-valued utilities and ensure existence of the largest utility for the most prefered alternative.
Utilities are therefore upper-bouned and disutilities, defined as negative utilities, are bounded from bottom.

The concept of utility dates back to the XVIII-th and XIX-th century utilitarian's economy and its eminent representative Jeremy Bentham.
In its further evolution it acquired quantitative and strict meaning, mainly due to Morgentern and von Neumann \cite{morgenstern}.
In order to use it in specific applications of choice models, one has to postulate a lot from outside of the theory and to find an effective way to either derive it from any fundamental or general theory or to parametrize it and estimate from data. For example, efficient estimates of utility in marketing is done in the framework of {\it conjoint} methods \cite{evgeniou}.
Derivation of the utility function from first principles represents major theoretical problem because of lack of fundamental theory of human behaviour.
Normally, one has to rely on partially justified models using power series approximations for utility functions, often restricted to linear or quadratic terms (cf. ref. \cite{dobson}), and finding empirically relevant variables and estimators for parameters, if any data are available at all.
For a broader discussion we refer to e.g. ref. \cite{hruszka}.

Apart from those problems, numerous conceptual difficulties arise when applying utility functions to quantify the behaviour of the choice makers and to predict them.
In particular, classical Bentham's approach of maximizing the overall utility, integrated over individuals, when taking market decisions, is often a subject of serious objections \cite{straffin}.

For each individual, utility is reduced whenever choice is performed, similarly to the wave-packet reduction in the process of quantum-mechanical measurement.
Consequently in this paper we denote utility as a random variable by $U$ and its particular instance, e.g. after making a choice, by $u$.
For disutilities we adopt notation $\bar U$ and $\bar u$, respectively.

The system consists of $N=\sum_{k=1}^M N_k$ agents, where $N_k$ stands for the number of agents on the $k$-th market.
Individuals' disutilities $\bar{U}_{i_k}=-U_{i_k}$ contribute additively to the overall network's disutility $\bar{U}=\sum_{i_k=1}^{V_k}\bar{U}_{i_k}$.
If markets are assumed to be atomic subsystems and agents do not interact then additivity of the overall system's disutility is ensured by construction \cite{majka1}. 
Indeed, decompose the system $\mathcal S$ into the sum of subsystems (markets) $\mathcal{S}_k$, $\mathcal{S}=\mathcal{S}_1\oplus\ldots\oplus\mathcal{S}_M$.
Then, by construction, additivity condition is fulfiled $\bar{U}=\bar{U}_1+\ldots +\bar{U}_M$ (for more detailed discussion of additivity cf. ref. \cite{touchette}).
Furthermore, for $\bar{U}=\mathcal{O}(N^{\alpha})$, $\alpha\le 1$, disutility is also extensive.

In order to follow classical Boltzmann arguments leading to the exponential probability $\mathcal{P}(\bar{U}=\bar u)\sim \exp(-\beta\bar{u})$, besides additivity we also assume conservation of disutility for the whole system.
Validity of this assumption is traditionally matter of debate (cf. e.g. \cite{luce}).
In economy one normally does not contest it for utility being function of money or any other conserved commodity.

We further assume existence of the stationary regime and existence of the first moment of disutility, $\la\bar U\ra <\infty$, at least in the limit of long times.
This ensures existence of the equilibrum temperature $T_{eq}=1/\beta_{eq}$ in that regime.

We consider the canonical and the grand canonical ensembles of systems (cf. ref. \cite{majka1}).
In the canonical ensemble the total number of agents $N$ is fixed and the partition function can be written as
\be
Z(\beta) = \prod_{k=1}^M Z_k^1(\beta)^{N_k},
\label{eq0a}
\ee
where $Z_k^1(\beta)$ stands for the partition function for the $k$-th market with one agent on it and $N_k=0,1,\ldots,\infty$.
Eqn. (\ref{eq0a}) assumes {\it non-subjectivity} of the utility, i.e. that utility of given route is the same for any decision maker.
For the grand canonical ensemble, or random $N$, one introduce another Lagrange multiplier, called the chemical potential $\mu$, and the grand partition function reads
\be
\Xi(\beta,\mu) = \prod_{k=1}^M \frac{1}{1-X_k},
\label{eq0b}
\ee
where $X_k=e^{\nu}Z_k^1(\beta)$, $\nu=\beta\mu$.
Moments of extensive variables and the entropy are given by the partition functions $Z$ and $\Xi$, and their derivatives over $\beta$ and $\nu$. 

In order to make thermodynamic description complete, we define the volume $V_k$ of the $k$-th market as the total number of choice alternatives available for each agent on this market, and the total volume of the system as $V=\sum_{k=1}^M V_k$.
Such defined total volume is additive by construction.
Using market volumes, partition functions (\ref{eq0a},\ref{eq0b}) can be written as
\be
\ln Z(\beta;\vec{N},\vec{V};M)=\sum_{k=1}^MN_k\ln\sum_{j_k=1}^{V_k}e^{-\beta\bar{u}_{j_k}}
\label{eq0c}
\ee
and
\be
\ln \Xi(\beta,\nu;\vec{N},\vec{V};M)=-\sum_{k=1}^M\ln(1-e^{\nu}\sum_{j=1}^{V_k}e^{-\beta\bar{u}_{j_k}}),
\label{eq0d}
\ee
where $\bar{u}_{j_k}$ stands for disutility of the $j_k$-th agent's choice on the $k$-th market, $\vec{N}=(N_1,\ldots,N_M)$, $\vec{V}=(V_1,\ldots,V_M)$.

We define the pressure as
\be
p=\frac{1}{\beta}\Big(\frac{\partial}{\partial V}\ln Z(\beta,V)\Big)_{\beta}
\label{eq01}
\ee
for the canonical ensemble and
\be
p=\frac{1}{\beta}\Big(\frac{\partial}{\partial V}\ln \Xi(\beta,\nu,V)\Big)_{\beta,\nu}
\label{eq001}
\ee
for the grand canonical ensemble.
Such defined pressure is naturally interpretable thermodynamically as an effect of the size of the choice set on the free utility $F$
\be
p=-\Big(\frac{\partial F}{\partial V}\Big)_{\beta (,\nu)}.
\label{eq002}
\ee

Since volumes are discrete variables, differentiation over them has to be defined with care.
We adopt an approach of the finite-difference calculus for functions defined on an equally-distant grid with a unit space.
Approximations for the $m$-th order derivatives operators can be expressed using the forward (backward)-difference operators $\Delta$ ($\nabla$)
\be
\frac{\partial^m}{\partial x^m} & = & \Big(\sum_{n=1}^{\infty}\frac{(-1)^{n-1}}{n}\Delta^n\Big)^m \nn \\
                                & = & \Big(\sum_{n=1}^{\infty}\frac{1}{n}\nabla^n\Big)^m,
\label{eq02}
\ee
where for any test function $f(x)$ ($x\in\mathbb{Z}_1$), $\Delta f(x)=f(x+1)-f(x)$ and $\nabla f(x)=f(x)-f(x-1)$.
Using the forward- or the backward-difference operators expansion depends on which one gives more accurate approximation to the derivative's values and this normally depends on the distance of $x$ from the domain's edge.

\section{Partition functions and thermodynamic variables}
\label{sec:2}
\subsection{The cumulant expansion}
\label{sec:2.1}

Eqns (\ref{eq01},\ref{eq001}) express the pressure $p$ via $\beta$, $\nu$ and volume $V$, and lead directly to equations of state.
Depending on the temperature regime, one can expand partition functions around any characteristic temperature $\beta_0$ and fugacity $\nu_0$ of the system.
The $\beta_0$ and $\nu_0$ are determined by the mean utility and number of agents.

The $\ln Z$ and $\ln\Xi$ are generating functions for the cumulants of $\bar{U}$ and $N$ \cite{abramowitz}.
The $\ln Z_k^1$ can be expanded as
\be
\ln Z_k^1(\beta,\beta_0,V_k)=\sum_{n=0}^{\infty}\lla\bar{U}_k^n(\beta=\beta_0,V_k)\rra\frac{(\beta_0-\beta)^n}{n!},
\label{eq03}
\ee
where the cumulants of disutility can be related to ordinary moments (cf. e.g. \cite{gardiner}), e.g.
\be
\lla\bar{U}\rra & = & \langle\bar{U}\rangle, \nn \\
\lla\bar{U}^2\rra & = & \langle\bar{U}^2\rangle-\langle\bar{U}\rangle^2, \nn \\
\lla\bar{U}^3\rra & = & \langle\bar{U}^3\rangle-3\langle\bar{U}^2\rangle\langle\bar{U}\rangle+2\langle\bar{U}\rangle^3, \nn \\
\ldots & & \ldots\;\;\mbox{etc.}
\label{eqxi04}
\ee 
For the grand canonical ensamble, expansion of $\Xi_k$ reads 
\be
\lefteqn{\ln \Xi_k(\beta,\beta_0,\nu,\nu_0,V_k) = }\nn \\
& & \sum_{n,m=0}^{\infty}\lla\bar{U}_k^n(\beta=\beta_0,\nu=\nu_0,V_k)\times \nn \\
& & N_k^m(\beta=\beta_0,\nu=\nu_0,V_k)\rra\times \nn \\
& & \frac{(\beta_0-\beta)^n}{n!}\frac{(\nu-\nu_0)^m}{m!}, 
\label{eq06}
\ee 
where the bivariate cumulants can be related to bivariate moments \cite{gardiner}, e.g.
\be
\lla\bar{U}N^1\rra & = & \la\bar{U}N\ra-\la\bar{U}\ra\la N\ra, \nn \\
\lla\bar{U}^2N^1\rra & = & 2\la\bar{U}\ra\la N\ra^2-2\la N\ra\la\bar{U}N\ra-\la\bar{U}\ra\la N^2\ra \nn \\
                     & + & \la\bar{U}^2 N\ra, \nn \\
\ldots & & \ldots \;\;\mbox{etc.}
\ee

\subsection{The high-temperature regime}
\label{sec:2.2}

Applying the cumulant expansion around $\beta=1/T=0$ for fixed $\nu$ and calculating the pressure (\ref{eq001}) we get
\be
p & = & \sum_{k=1}^M\frac{\pa V_k}{\pa V}\Big(\frac{1}{\beta}\frac{\la N_k\ra}{V_k}-\frac{\pa\la\bar{U}\ra}{\pa V_k}\Big|_{\beta=0}+\frac{\beta}{2}\frac{\pa\,\mbox{Var}(\bar{U})}{\pa V_k}\Big|_{\beta=0} \Big) \nn \\
  & + & \frac{1}{\beta}\Big(\frac{\pa\ln\Xi}{\pa M}\Big)_{\beta,\nu,\vec{N}}\frac{\pa M}{\pa V}.
\label{eq07}
\ee
For $M=1$ the first term in eqn (\ref{eq07}) corresponds to the equation of state of the ideal classical gas: $pV\beta=\la N\ra$.
For fixed $M$ also $\pa M/\pa V=0$.
Corrections to the ideal case depend on the network topology and the utility function.

\subsection{Temperature evolution of the pressure}
\label{sec:2.3}

As found in ref. \cite{majka2}, using R\'enyi entropies, defined for any $q\in\mathbb{R}$ ($J$ standing for number of states)
\be
I_q & = & \left\{ \begin{array}{cc}
          \frac{1}{1-q}\ln\sum_{j=1}^J p_j^q, & \;\;\;\;\;q\ne 1 \\
          -\sum_{j=1}^J p_j\ln p_j, & \;\;\;\;\;q=1
          \end{array}\right .
\label{eq071}
\ee
one relates partition functions at different temperatures, with $q$ being scaling parameter
\be
\ln\Xi(q\beta,q\nu,V)-q\ln\Xi(\beta,\nu,V)=(1-q)I_q(\beta,\nu,V).
\label{eq072}
\ee
and similarly for $Z$.

Using eqn. (\ref{eq01}), this leads to more general equation
\be
p(q\beta,q\nu)=qp(\beta,\nu)+\frac{1}{\beta}(1-q)\frac{\pa I_q(\beta,\nu,V)}{\pa V},
\label{eq073}
\ee
being not only equation of state but determining also the temperature evolution of the pressure.

\section{Applications}
\label{sec:3}
\subsection{Networks of increasing path lengths: analogues to quantum-mechanical systems}
\label{sec:3.1}

Assuming disutility function depending only linearly or quadratically on the topological $(OD)$ distance, for Hamilton networks containing routes of all lengths one observes direct analogies to quantum-mechanical systems: quantum harmonic oscillator and the ideal quantum gas. 

\begin{figure}[h]
\begin{center}
\begin{tabular}{cc}
\includegraphics[scale=.3]{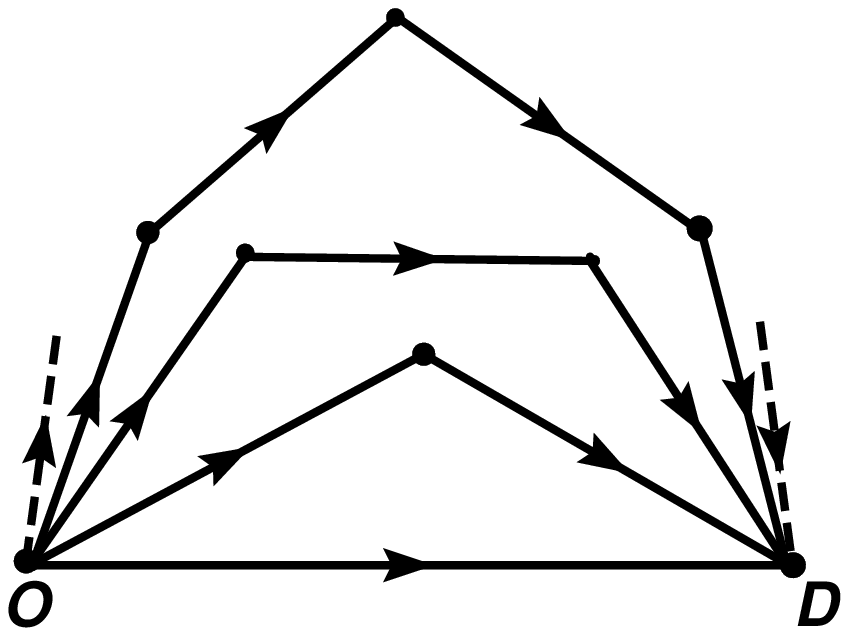} & \hspace{5mm} \includegraphics[scale=.3]{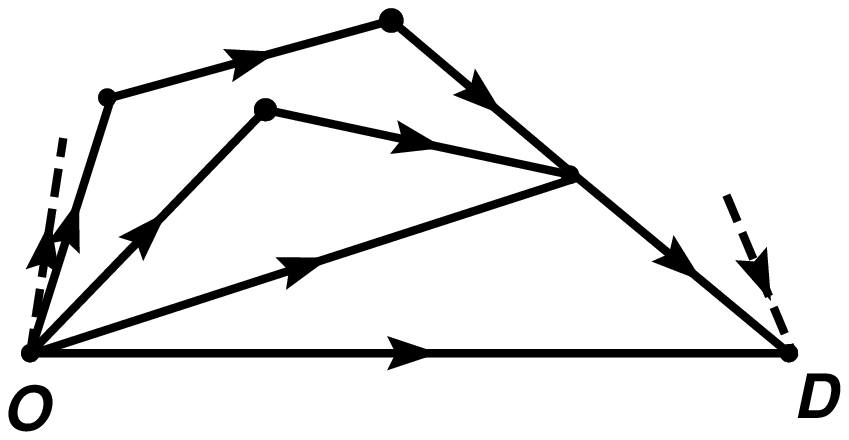} \\
\includegraphics[scale=.3]{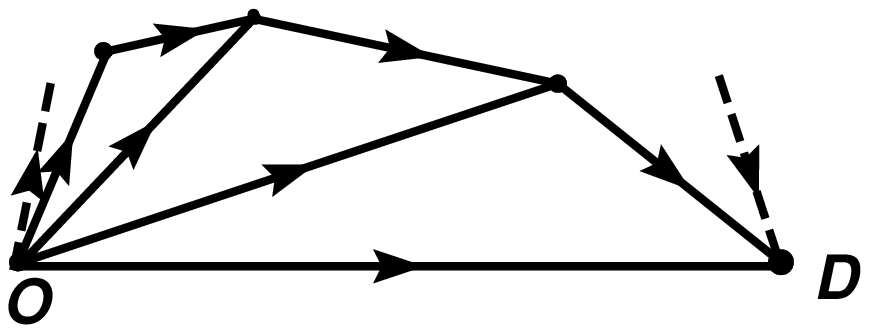} & \hspace{5mm} \includegraphics[scale=.3]{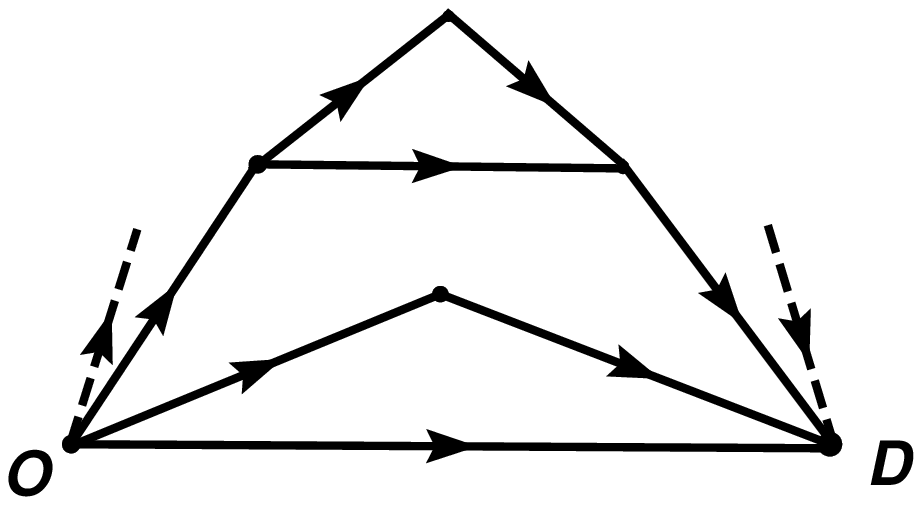}
\end{tabular}
\caption{\label{fig01}\em Example network topologies with one market and route lengths increasing linearly in steps of one.}
\end{center}
\end{figure}

\begin{figure}[h]                                                  
\begin{center}
\begin{tabular}{cc}
\includegraphics[scale=.3]{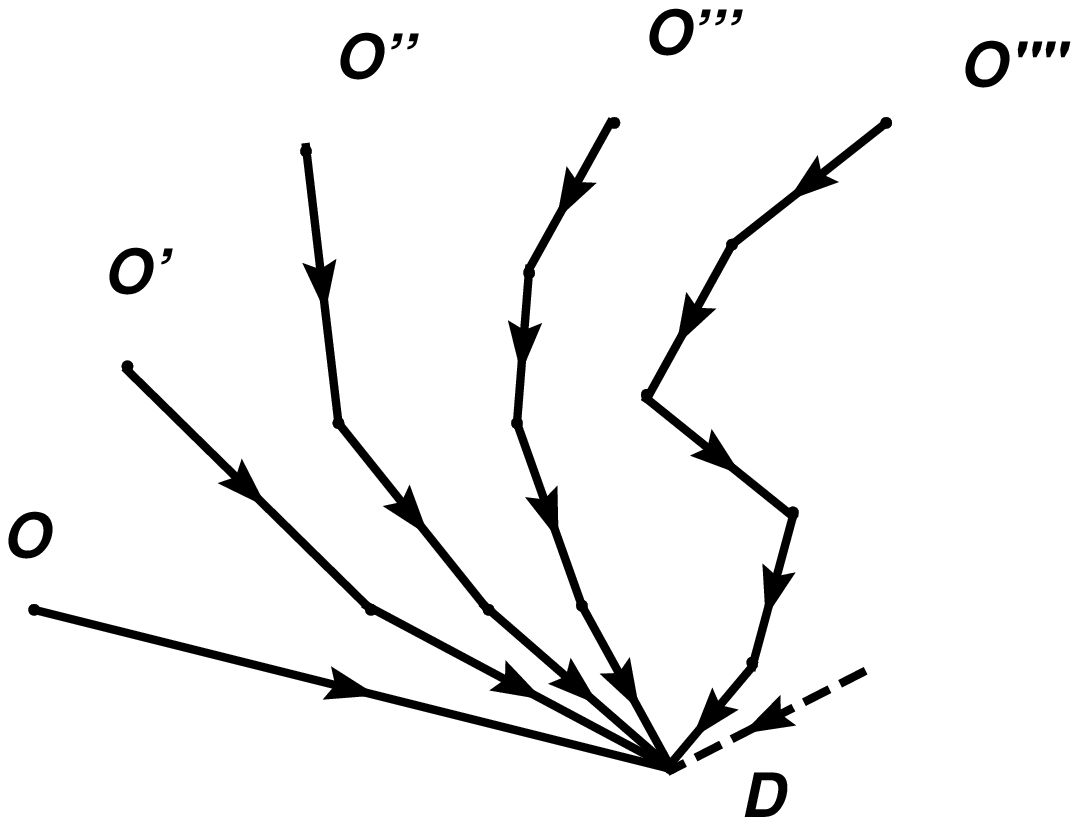} & \includegraphics[scale=.3]{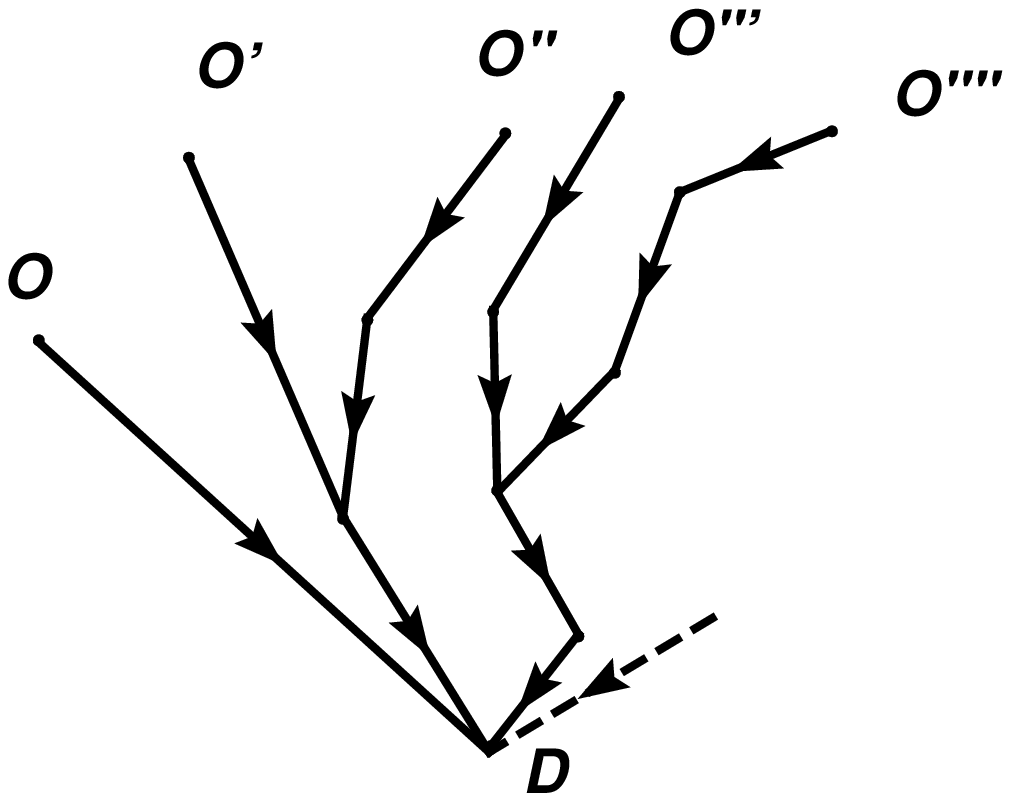} \\           
\includegraphics[scale=.3]{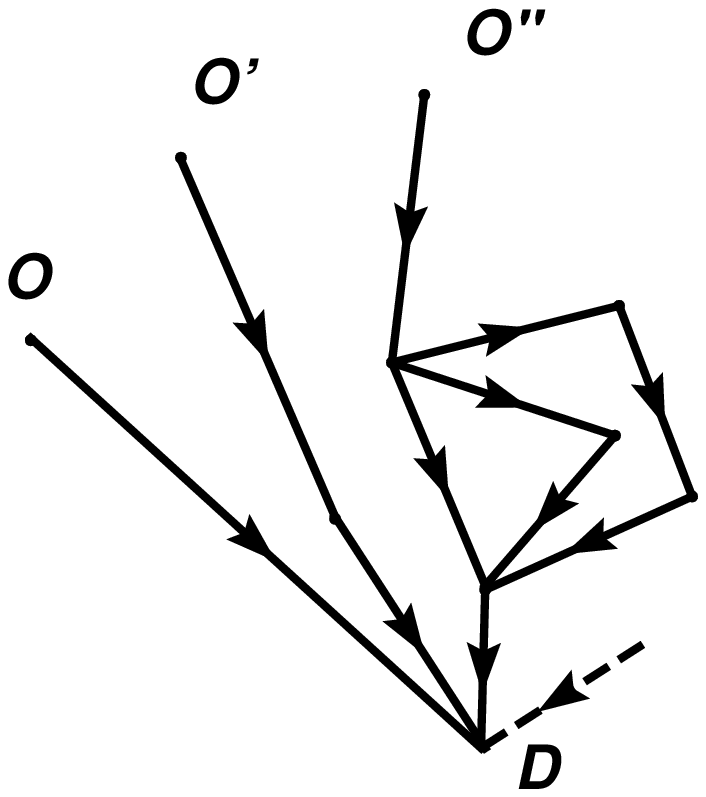} & \includegraphics[scale=.3]{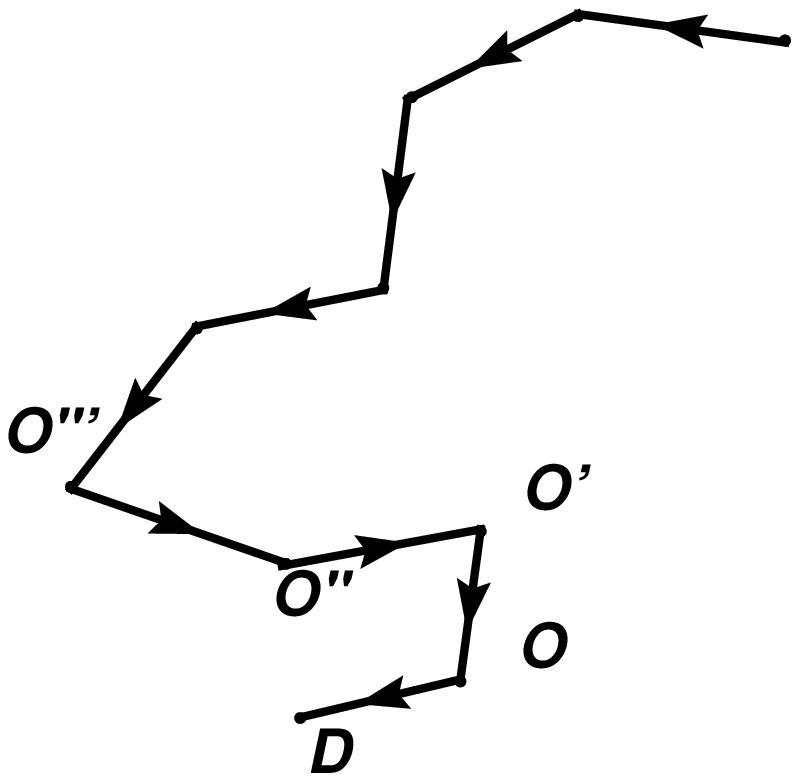}              
\end{tabular}
\caption{\label{fig02}\em Example topologies for multi-market networks of routes with lengths increasing linearly in steps of one.}
\end{center}                                                                                     \end{figure}

Networks containing routes of all lengths may consist of one market (one $(OD)$ pair) or many markets (more than one $(OD)$ pair).
Such classes of networks are very broad, even excluding non-compact and tadpole graphs.
Examples of one-market and multi-market networks are given in Figs.~\ref{fig01} and \ref{fig02}, respectively. 
Dependence of the market volume $V$ on the number of nodes $L$ is $\sqrt{L}$ (upper left in Fig.~\ref{fig01}), linear for one-market networks (lower left in Fig.~\ref{fig01}), and quadratic for multi-market networks (upper left in Fig.~\ref{fig02}) which makes these networks rather easy to simulate.

Assume agent's disutility proportional to the number $n$ of route segments, $\bar{U}_n=\bar{u}_0+n\bar{u}_1$ ($n=0,1,\ldots$), where $\bar{u}_{0,1}$ are constants.
For one agent, this model exhibits analogy to the one-dimensional quantum harmonic oscillator with the energy spectrum $E_n=\hbar\omega(1/2+n)$.
For $N$ agents, additivity of disutility implies $\bar{U}_{\vec n}=\bar{u}_0+\bar{u}_1\sum_{i=1}^N n_i$, $\vec n=(n_1,\ldots,n_N)$, and is the same as the energy spectrum of the $N$-dimensional oscillator $E_{\vec n}=\hbar\omega(N/2+n_1+\ldots+n_N)$.
The values of constants $\bar{u}_0=0,\; \bar{u}_1=1$ are assumed without loss of generality.

For any oscillator-like topology, excluding $n=0$ routes corresponding to tadpole graphs, the statistical sum reads
\be
\lim_{V\rightarrow\infty}Z(\beta,V) & = & \lim_{V\rightarrow\infty}\Big(\frac{1-e^{-\beta V}}{e^{\beta}-1}\Big)^N \nn \\
                                    & = & \Big(\frac{1}{e^{\beta}-1}\Big)^N
\label{eq08}
\ee
for $\beta>0$.
Using first 3 terms of the expansion (\ref{eq03}) one gets the equation of state
\be
p=\frac{N}{\beta V}-\frac{N}{2}+\frac{N}{12}\beta V.
\label{eq09}
\ee
The first term of eqn. (\ref{eq09}) corresponds to the ideal classical gas and it dominates at high temperature.
The sign of correction to the ideal case is positive or negative, for $\beta V\ge 6$ or $\beta V< 6$, respectively, for which we hardly see any analogy in gas dynamics.
At the equilibrium temperature $\beta=\ln [\bar{U}/(\bar{U}-N)]$ the pressure is equal to
\be
p & = & \frac{N}{\bar{U}/(\bar{U}-N)^V-1} \nn \\
  & \sim & N\Big(1-\frac{N}{\bar{U}}\Big)^V\;\;\;\;\;\;\;\mbox{for $V\gg 1$}.
\ee 
Considering random number of agents $N$, at the equilibrium temperature $\beta=\ln [\la\bar U\ra/(\la\bar U\ra-\la N\ra)]$ and equilibrium fugacity $\nu=\ln [\la N\ra^2/(\la\bar{U}\ra-\la N\ra)(1+\la N\ra)]$, the equation of state reads:
\be
p = \frac{\la N\ra}{[\la\bar{U}\ra/(\la\bar{U}\ra-\la N\ra)]^V-1}.
\label{eq10}
\ee 

Let us assume agent's disutility proportional to the square of the route length, $\bar{U}_n=\bar{u}_2n^2$.
For the topologies of Fig. \ref{fig02} and for one agent this model corresponds to the ideal quantum gas in a one-dimensional box and for the $N$-agent disutility $\bar{U}_{\vec{n}}=\bar{u}_2\sum_{i=1}^Nn_i^2$, the analogy extends to the $N$-dimensional box.
Depending on the number of agents allowed on each market, $N=0,1,\ldots,\infty$ or $N=0,1$ (Pauli principle) and taking $\bar{u}_2=1$, one reproduces the case of the Bose-Einstein or Fermi-Dirac statistics, respectively, for which the grand partition functions are (cf. e.g. ref. \cite{balescu})
\be
\ln\Xi(\beta,\nu) = \left\{ \begin{array}{ll}
                          -\sum_{\vec{n}} \ln(1-e^{\nu}e^{-\beta \bar{U}_{\vec{n}}}), & \;\;\;\mbox{\small bosons} \\
                           \sum_{\vec{n}} \ln(1+e^{\nu}e^{-\beta \bar{U}_{\vec{n}}}), & \;\;\;\mbox{\small fermions.}
                           \end{array} \right .
\label{eq11}
\ee
The Fermi-Dirac case can be interpreted as treating the whole group of agents as a single decision maker acting on the market ($N=1$) or remaining passive ($N=0$).
More general dependence of $\bar{U}$ on $n$ may also correspond to non-ideal quantum systems or different boundary conditions, or topology different than in Fig. \ref{fig02}.

Formulae for R\'enyi entropies of quantum gases are given in refs. \cite{majka2} and can be almost directly used for the evolution equation (\ref{eq073}).
Explicit summation over states has to be performed, depending on the network topology (cf. Fig.~\ref{fig02}).

\subsection{Maximum-connectivity networks}
\label{sec:3.2}

\begin{figure}[h]
\begin{center}
\includegraphics[scale=.35]{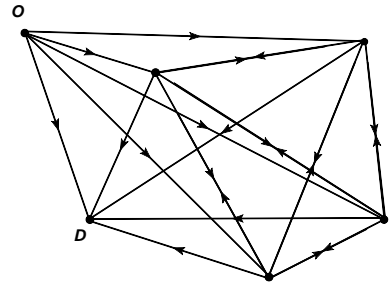} 
\caption{\label{fig03}\em Maximum-connectivity network.}
\end{center}
\end{figure}
For the {\it maximum-connectivity} or {\it complete} network (cf. Fig. \ref{fig03}) all $M=L(L-1)$ markets are topologically identical.
Within linear utility approximation one finds
\be
Z_k^1(\beta)=\sum_{l=1}^{L-1}\frac{(L-2)!}{(L-1-l)!}e^{-\beta\bar{u}_1 l}
\label{eq12}
\ee
and 
\be
V_k=(L-2)!\sum_{l=0}^{L-2}\frac{1}{l!}.
\ee
For $L\gg 1$ and $\bar{u}_1=1$ one approximates $V=MV_k\simeq eL^2L!$ and, keeping only leading terms in $L$, arrives to the equation of state
\be
p=\left\{\begin{array}{cc}
  \frac{N}{\beta V}-\frac{N}{V\ln L}, & \;\;\;\;\; L\ge e^{\beta} \\
  0,                                  & \;\;\;\;\; L< e^{\beta}.
         \end{array}\right .
\label{eq13}
\ee  
For high temperature, $\beta\le\ln L$, and lowly populated network, $N\ll V\ln L$, the equation of state of the ideal clasical gas is recovered from eqn. (\ref{eq13}).
Correction to the ideal case is always negative, as for attracting molecular forces and quantum Bose statistics.

\begin{figure}[h]
\begin{center}
\includegraphics[scale=.5]{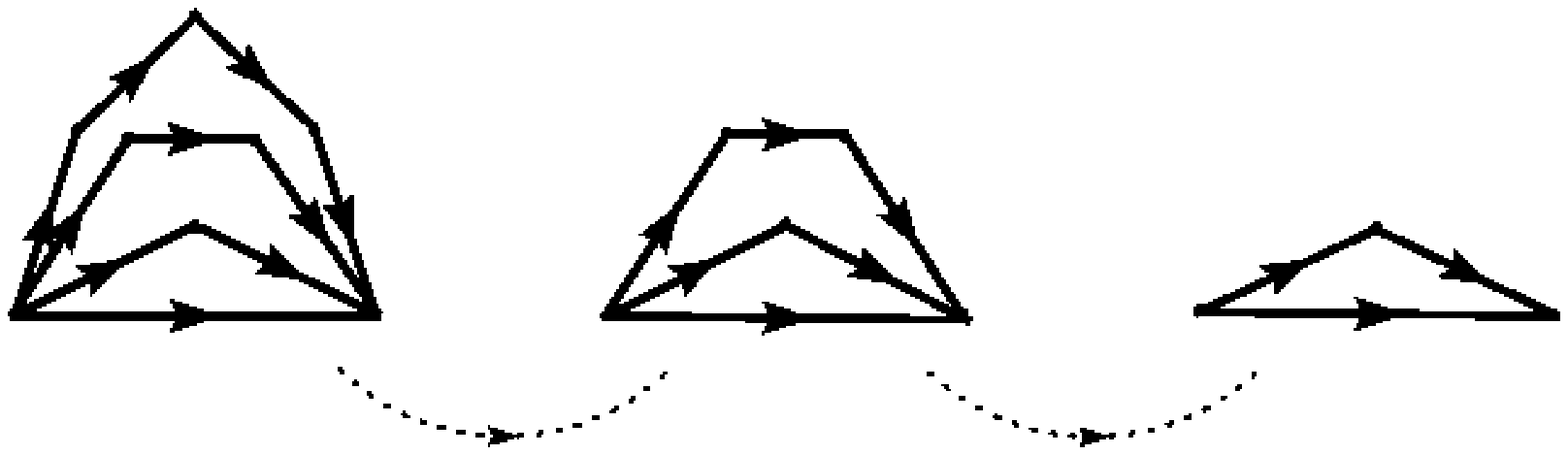} 
\vspace{0mm}
\includegraphics[scale=.5]{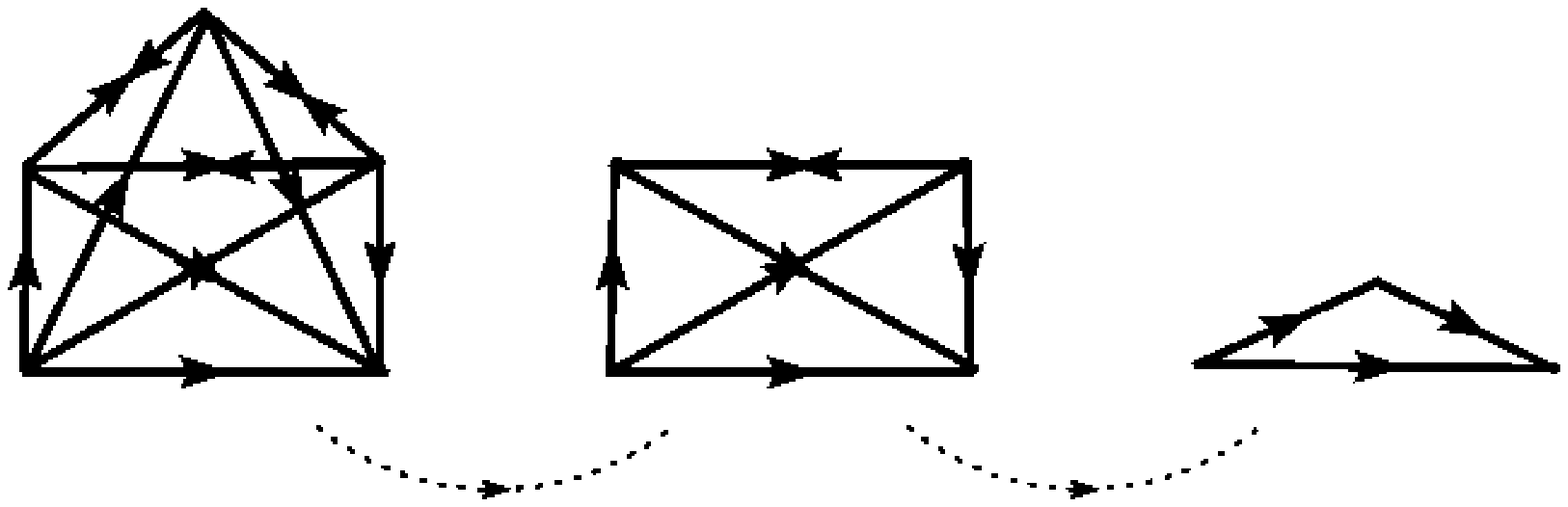}
\caption{\label{fig04}\em Two sequences of networks evolving towards triangle network by decreasing numbers of nodes and keeping their topological identity: maximum-connectivity network (lower) and harmonic oscillator network (upper).}
\end{center}
\end{figure}
In general, equation of state for high temperature can be derived from eqn. (\ref{eq07}).
For fixed number of markets, i.e. the number of markets independent of size of the choice set, $\pa M/\pa V=0$, the formula for ideal gas is recovered.
For variable $M$ we find the volume $V=eL(L-1)\Gamma (L-1,1)$, where $\Gamma (.,.)$ stands for incomplete Euler gamma function \cite{abramowitz}, and accounting only for terms at $1/\beta$ in expansion (\ref{eq07}) we get
\be
p=\frac{N}{\beta V}+\frac{N}{\beta e(2L-1)\Gamma^2(L-1,1)}\frac{\pa\Gamma (L-1,1)}{\pa L}.
\label{eq14}
\ee
Second term in (\ref{eq14}) represents the finite-$L$ correction to the ideal-gas term $N/\beta V$ and it vanishes for $L\rightarrow\infty$, in agreement with (\ref{eq13}).

Interesting property of network dynamics can be illustrated when calculating pressure in the simplest non-trivial case of the complete $L=3$ graph, being at the same time the oscillator-like network with two routes, $V=2$.
The pressure depends on the model we choose: keeping terms proportional to the $0$-th and $-1$-st powers of $\beta$ in eqs. (\ref{eq09}) and (\ref{eq13}) one obtains two different solutions
\be
p=\left \{\begin{array}{cc}
          \frac{N}{\beta V}-\frac{N}{4}, & \;\;\;\;\;\mbox{\small maximum-connectivity} \\
          \frac{N}{\beta V}-\frac{N}{2}, & \;\;\;\;\;\mbox{\small harmonic oscillator.}
          \end{array}\right .
\label{eq15}
\ee
This is understandable because the pressure depends on the change of utility with volume and not on the volume itself.
Similar behaviour we observe in a physical system where the pressure depends on the way the volume was changed.
Moreover, the pressure of physical gas depends on its statistical properties, e.g. ideal fermions exhibit higher pressure than bosons in the same conditions.
In our case network topology affects the properties of the system.
Two ways of evolution of network topology leading to the same final state are shown in Fig.~\ref{fig04}.

\subsection{Hub-and-spoke networks}
\label{sec:3.3}
                                                                                                 \begin{figure}[h]
\begin{center}
\includegraphics[scale=.35]{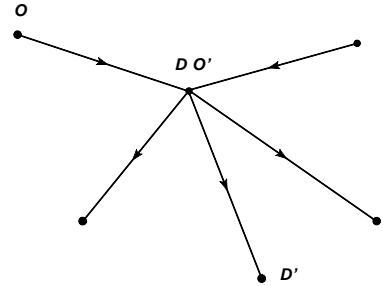}
\caption{\label{fig05}\em Hub-and-spoke network.}
\end{center}
\end{figure}
For {\it hub-and-spoke} networks, being often met in local airline transportation \cite{dobson} or resource brokerage in Grid computing \cite{crosby}, there are only two classes of topologically different markets: $M_1=L^2-3L+2$ spoke-spoke and $M_2=2(L-1)$ spoke-hub, as shown in Fig.~\ref{fig05}.

The partition function
\be
Z(\beta)=\prod_{k_1=1}^{M_1}e^{-2\beta\bar{u}_1 N_{k_1}}\prod_{k_2=1}^{M_2}e^{-\beta\bar{u}_1 N_{k_2}},
\label{eq16}
\ee
where $N_{1(2)}$ are agents' populations on both types of markets, does not depend on the volume $V=L(L-1)$ and thus the pressure is zero.
However, the grand partition function $(\bar{u}_1=1)$
\be
\Xi(\beta,\nu)=\frac{1}{(1-e^{\nu-2\beta})^{M_1}}\frac{1}{(1-e^{\nu-\beta})^{M_2}}
\label{eq17}
\ee
does depend on the volume and leads to interesting equation of state.
The complete formula is very lengthy and we quote here only its large-volume limit
\be
p=\frac{\la N\ra}{\beta V}+\frac{\la N\ra(1-e^{\beta})}{\beta V^{3/2}},
\label{eq18}
\ee
reducing further in the high-temperature regime to
\be
p=\frac{\la N\ra}{\beta V}-\frac{\la N\ra}{V^{3/2}}.
\label{eq19}
\ee
The same result can be obtained directly from eqn. (\ref{eq07}).
For $\la N\ra\ll V^{3/2}$ formula (\ref{eq19}) describes ideal classical gas.
Positivity of the pressure $(V>\beta^2)$ is ensured in this approximation because $\beta^2\ll1$ and $V>1$. 

\section{Discussion and final remarks}
\label{sec:4}

Systems of choice makers on networks can be treated thermodynamically, using utility function in a way analogous to the energy function and defining the system volume as the number of choice alternatives. 
Working within the framework of equilibrium thermodynamics we found equations of state and discussed them in detail for specific network topologies, utility functions and in different temperature regimes.

For a single-market and multi-market networks with linear utilities and route lengths covering all integer numbers with no degenaracy, close resemblance to the one- or multi-dimensional quantum harmonic oscillator was found.
Equation of state, besides the ideal-gas terms, contains corrections with no clear analogy in gas dynamics.
For a quadratic utility function the system exhibits similarity to the ideal boson or fermion gas, depending on the number of agents allowed on one market.

In case of the maximum-connectivity network, apart from deriving equation of state, we also discussed its dependence on the evolution of the connectivity scheme with volume.
This property we found to be a particularly interesting example of how the network structure determines global properties of the system.
It may suggest a possibility of phase transitions driven by variation of network topology and, we believe, deserves further investigation.

We had a closer look at the hub-and-spoke connectivity scheme as being important for applications.
Analytical solutions for such system can be found and they simplify in the large-volume limit where corrections to the ideal-gas terms are negative.
They are exponential with $1/\beta$ quenching in inverse temperature and depend on the volume as $V^{-3/2}$.

In future work, more attention is certainly called for elucidating the role of possible correlations between agents, correlations between markets and of the form of utility function for this class of complex systems.
Any mechanism giving dependence between subsystems, either explicit correlation of choice alternatives \cite{sznajd} or interaction term between utilities \cite{durlauf}, or limited node or link transmitivity, may result with non-additivity of utility and eventually need for using non-extensive thermodynamics \cite{plastino}.
In particular, it is natural to expect a non-linear reinforcement of utility for the system of two agents making the same choice.
Investigation of correlations between markets, either coming from explicit interactions between agents or from network constraints, should proceed along two lines. 
First, for weak correlations between subsystems one stays in the framework of Boltzmann-Gibbs thermodynamics and accounts only for agent interactions.
Second, in case of unavoidable non-extensivity, the whole formalism has to be reworked using power-law probability distributions. 
It would be very interesting if any nontrivial modification of the topology-dependence of the equation of state is observed, as compared to found in this work.


\begin{thebibliography}{}
\bibitem{majka1} A. Majka and W. Wi\`slicki, Physica A337(2004)645.
\bibitem{majka2} W. Wi\`slicki, J. Phys. A34(2001)4663;
                 A.~Majka and W.~Wi\`slicki, Physica A322(2003)313.
\bibitem{burgos} E. Burgos et al., Phys. Rev. E65(2002)036711; 
                 J. Berg Phys. Rev. E61(1999)2327; 
                 J. Berg and M. Lassig, Phys. Rev. Lett. 89(2002)228701;
                 D. Challet et al., Phys. Rev. Lett. 84(2000)1824;
                 M. Woolf et al., Phys. Rev. E66(2002)046106;
                 P. de los Rios et al., Phys. Rev. E53(1996)R2029;
                 F. Seno et al., Phys. Rev. E55(1997)3859;
                 N. Schwartz et al., Phys. Rev. E58(1998)7642;
                 L.A. Braunstein et al., Phys. Rev. Lett. 91(2003)168701.
\bibitem{ducheneaut} N. Ducheneaut and L.A. Watts, Human-Comp. Interaction 20(2005)11.
\bibitem{bierlaire1} M. Bierlaire, {\it Discrete Choice Models}, in M. Labb\'e, G. Laporte, K. Tanczos and Ph. Toint (eds), {\it Operations Research and Decision Aid Methodologies in Traffic and Transportation Management}, NATO ASI Series, Series F: Computer and Systems Sciences, Vol. 166, Springer Verlag, pp. 203-227
\bibitem{morgenstern} J. von Neumann and O. Morgenstern, {\it Theory of Games and Economic Behaviour}, John Wiley, 1967 (original edition, 1944).
\bibitem{evgeniou} T. Evgeniou, C. Boussios and G. Zacharia, Marketing Science, 24(2005)415.
\bibitem{dobson} G. Dobson and F.J. Lederer, Transp. Science 27(1993)281.
\bibitem{hruszka} H. Hruszka, W. Fettes and M. Probst, Eur. J. Oper. Res. 159(2004)166.
\bibitem{straffin} P.D. Straffin, {\it Game Theory and Strategy}, The Mathematical Association of America, Washington D.C., 1993.
\bibitem{touchette} H. Touchette, Physica A305(2002)84.
\bibitem{luce} R.D. Luce and H. Raiffa, {\it Games and Decisions}, John Wiley, 1957
\bibitem{abramowitz} M. Abramowitz and I. Stegun, {\it Handbook of Mathematical Functions}, National Bureau of Standards, Applied Mathematics Series 55, 1972.
\bibitem{gardiner} C.W. Gardiner, {\it Handbook of Stochastic Methods}, Springer Series in Synergetics, Springer, 1986
\bibitem{balescu} R. Balescu, {\it Equilibrium and Nonequilibrium Statistical Mechanics}, A Wiley-Interscience Publication, John Wiley and Sons, 1975
\bibitem{crosby} P. Crosby, D. Colling and D. Waters, IEEE Trans. Nucl. Sci. 51(2004)884.
\bibitem{sznajd} K. Sznajd-Weron and J. Sznajd, Int. J. Mod. Phys. C11(2000)1157.
\bibitem{durlauf} S.N. Durlauf, {\it Proceedings of the National Academy of Sciences}, 96(1999)10582;
                  W. Brock and S.N. Durlauf, {\it Interaction-based models}, in {\it Handbook of Economics}, Vol. 5, Elsevier Science B.V., 2001, pp. 3297-3380.
\bibitem{plastino} A. Plastino, R. Mendes and C. Tsallis, Physica A261(1998)534.
\end{thebibliography}
\end{document}